# Cloud-Based Content Cooperation System to Assist Collaborative Learning Environment

Geunsik Lim[1], Donghwa Lee[2], and Sang-Bum Suh[3]
Software Center[123]
Samsung Electronics[123]
Suwon 443-370, Republic of Korea[123]
{geunsik.lim[1], dh09.lee[2], sbuk.suh[3]}@samsung.com

*Abstract*— Online educational systems running on smart devices have the advantage of allowing users to learn online regardless of the location of the users. In particular, data synchronization enables users to cooperate on contents in real time anywhere by sharing their files via cloud storage. However, users cannot collaborate by simultaneously modifying files that are shared with each other. In this paper, we propose a content collaboration method and a history management technique that are based on distributed system structure and can synchronize data shared in the cloud for multiple users and multiple devices.

## I. INTRODUCTION

Recently, many users have been able to engage with online education systems where they can collaborate with one another over long distances through their smart devices. "Smart devices" refers to the mobile devices that can download and install an application from an application store that corresponds to the specific smart device in use. Cloud-based online education enables students to learn using on-line materials, even though students do not physically attend school. However, existing cloud services [1] do not have a suitable synchronization facility for the case in which the system has to synchronize shared data among the users and when multiple users try to access the shared data at the same time. Also, existing cloud storage [2] is not aware of multiple users using multiple devices.

In other words, current systems cannot ensure the consistency of the shared data because only one user can modify the shared file. For example, *Dropbox* and *GoogleDrive* produce a transaction error, [3] called a *file conflict*, if the two users try to save a modified version of the same shared file. Moreover, existing cloud services focus on the case of a single user with multiple devices rather than the case where more than one user is collaborating. Therefore, we need to research next generation cloud storage techniques that can support network synchronization [4] for content collaboration among users within an internet-of-things (IoT) environment. IoT refers to the interconnection of uniquely identifiable embedded computing devices connected to the existing Internet infrastructure. IoT is expected to offer advanced connectivity to devices, systems, and services that goes beyond machine-to-machine communication.

The remainder of this paper is organized as follows. Section II gives an overview of related work. Section III addresses the design and implementation of the proposed techniques. Section IV shows the experimental results. Finally, Section V provides the conclusion for the paper.

## II. RELATED WORK

*Dropbox* [5] supports file sharing and network file synchronization in order to easily share files over the cloud anywhere. Cloud computing [6] helps user to share data and to synchronize such data among the user's devices. However, current systems do not support cases where users modify shared files simultaneously. Moreover, users cannot edit shared files at the same time without causing *file conflicts* for cases where there are multiple users with multiple devices.

*GoogleDrive* [7], [8] has complete support for different operating systems, such as Windows, Linux, and Android, and provides a document-based communication network to modify web-based shared documents. Moreover, many users can simultaneously modify shared files through the transaction mechanism provided by Google docs. However, *GoogleDrive* focuses on providing file sharing services using a public cloud rather than a private cloud. Also, this system does not integrate the history management of shared files and the cloud storage. In other words, there is no history about the particular changes.

This paper addresses the use case where a multi-user-aware file sharing technique and history management technique are implemented to share files used within collaborative learning and teaching environment. Our proposed system describes a private cloud [9] that can provide cloud storage for users.

## III. PROPOSAL: CLOUD-BASED CONTENT COOPERATION SYSTEM

Our proposed system manages changes in the content using a modified content unit without access to the time unit of the file in order to synchronize content. Moreover, the proposed system supports a synchronization mechanism [10] that is based on the spin-lock algorithm for the case where many users try to access the shared file. These approaches overcome the problem where the consistency of the shared file is broken due to the simultaneous access case of the shared file for environments with multiple users and multiple devices.

### A. Design

We describe how to interconnect distributed revision control and cloud computing in order for many users to be able to edit the same content in cloud storage. This paper focuses on a cloud-based content collaboration system that can automatically merge and track shared content in real time.

The proposed system only requires a cloud server without a source code management server because our system is based on the synchronization of distributed cloud files. [11] In other words, we do not need a source code management (SCM) server to manage the modified content of a shared file. We can synchronize the data that has changed by using the cloud computing infrastructure and implementing the history management of the file with distributed clients. This technique has been named the "cloud-based content collaborative system to assist in a collaborative learning and teaching environment" (COCO).

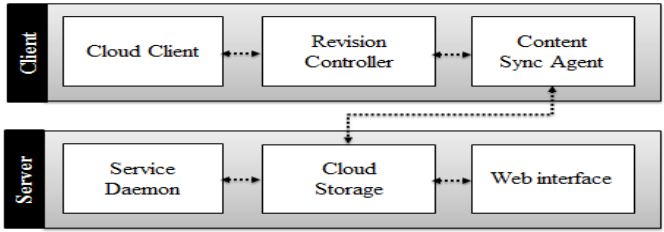

**Figure 1. Operation flow of the proposed system (COCO)**

Figure 1 shows the entire process flow of the proposed system. The proposed system consists of server-side and client-side. The server-side synchronizes and archives the user data in the cloud server, and the server side consists of three components as follows:

- Service daemon: synchronizes the shared data when the users modify the data held in the cloud server.
- Cloud storage: saves the content generated by the users.
- Web interface: supports file upload and download via a web browser.

The client side transfers the status and the content of the data to server side when the users modify the contents of the file. It consists of three components as follows:

- Content sync agent: synchronizes the content between the server and the client in real time.
- Revision controller: manages the changed file content using a distributed client approach.
- Cloud client: manages the content and information of the client.

In this paper, we contribute three achievements that are useful for an online collaborative education system in real time as follows:

- Changed data-based content synchronization: manages the changes in the file content without using a file access unit to support fine-grained synchronization function.
- Distributed content tracking: supports the distributed system mechanism [12] using a peer-to-peer technique without a centralized model.
- Automatic revision control: considers an environment with multiple users and multiple devices in order to solve the *file conflicts* that happen in a case where many users try to modify shared content.

*B. Changed Data-Based Content Synchronization*

The existing synchronization technique checks the access status of the data against the file access time. In this case, when two users try to simultaneously save the shared file, the users have to manually manage the conflicts in files due to a network failure.

To solve this problem, we have to avoid opening the shared file in case that another user opens a shared file that has already been opened by another user. When the user tries to close the opened file after modifying it, the client side must operate using an unlocking mechanism due to the locked status of the shared file. Next, the client has to guarantee the status of the file to enter the locking operation when the user of another device tries to modify the shared file.

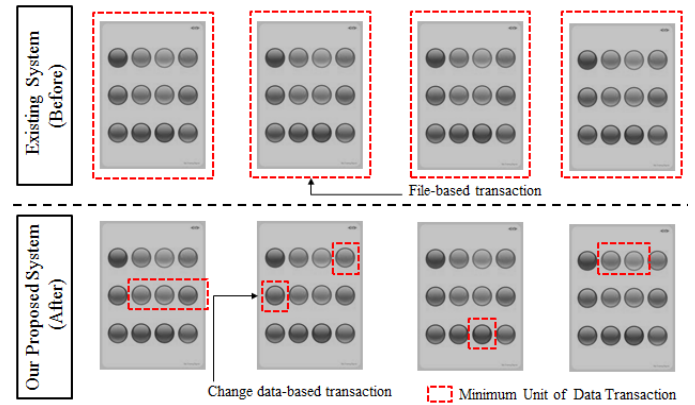

**Figure 2. Operation comparison of data synchronization between the existing system and the proposed system**

At this time, the server side preserves the information of the modified content in case that the shared file is modified by another user. It is possible for us to easily manage collaboration by using the history of the file content against the preserved information. Figure 2 shows how we can control the history management of files using the modified file content as the transaction's minimal unit, as compared to that of the existing system. The existing system cannot handle modifications of the same file in order to avoid file conflicts because it performs a transaction [9] for the file modification by using the access time to synchronize the shared file in the cloud without using any locking mechanism for the cloud file. Our proposed system solves the problem of *file conflicts* of the existing system when many users try to modify the content of the shared file with the modified content based locking mechanism.

*C. Distributed Content Tracking*

Figure 3 shows the distributed content control structure that can manage information of the changed file content when users modify the shared cloud file at the client side. The cloud based distributed history management technique does not need management system that manage the changed history of the clients. The history information of modified contents from the client can be replaced with cloud based network synchronization technique.

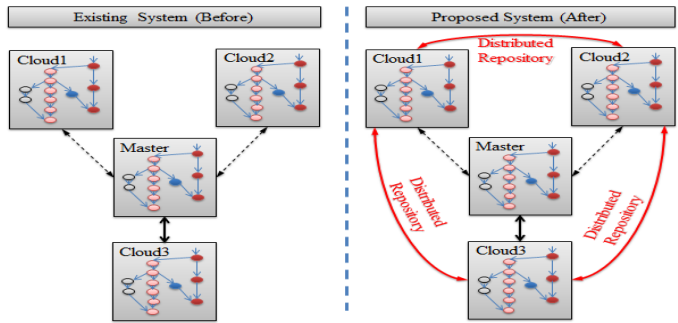

**Figure 3. The difference with respect to content history management between the existing system and the proposed system. The proposed system does not need to have single central place (e.g. *master*) where users commit their changes. Each user can have its own repository, and he can pull from each other repositories.**

*D. Automatic Revision Control*

Figure 4 shows how the proposed system synchronizes data between the server-side and the client-side in the case where a user creates new content using a mobile device so that the system can recognize a multi-user and multi-device environment. [13]

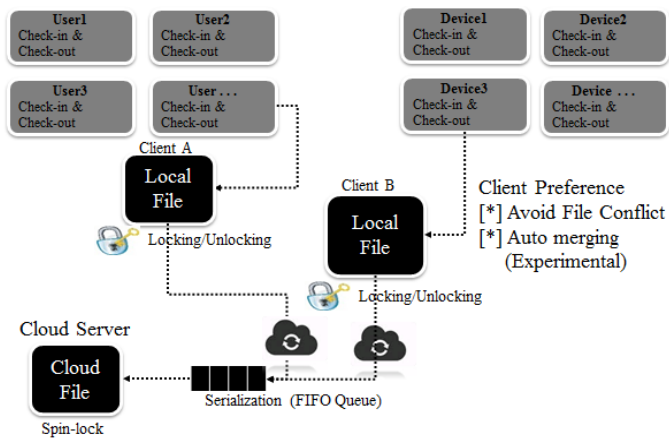

**Figure 4. Flow chart for content collaboration among the users**

The proposed system synchronizes data by processing the *check-in* and *check-out* according to the access status of the user's shared file. The client automatically merges the user data when another client sends the *check-in* commands to the cloud server. With current systems, if changes are automatically merged whenever the user tries to save modified content of a shared file, too much network overhead can be generated. Therefore, if we try to close the modified file when the client manually sends the *check-in* command to the cloud server, we can minimize the network overhead for frequent *send/receive* operations when the content of the shared file changes. At this time, the cloud server serializes the requests from the clients using a *First in, First out* (FIFO) behavior. [14] This technique is effective because the user can limit network usage over 3G networks every month. [15]

Figure 5 explains how the proposed system shares the data in the cloud storage while also managing the revision control system when the user modifies the shared file. At this time, a history management repository that integrates the modified shared data [16] is a client without a server. Each client is allocated a unique directory with a different name. [17] When the file content is modified using a directory that exists with a different name, the client automatically synchronizes the modified content to cloud server. Next, all of the clients' shared directory keeps the same file and same history information by synchronizing the network the same as with a file in the cloud of the different device using the most up-to-date data standard that is updated in the cloud server. The cloud server executes a serialization when the user saves the content of the file in the cloud according to the time of the cloud server.

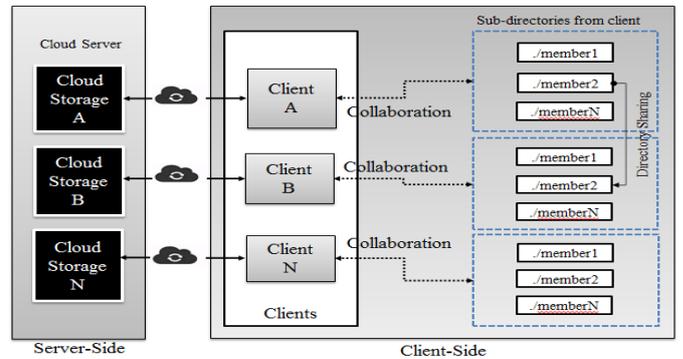

**Figure 5. The merging flow of the distributed cloud file for collaborative content history management**

IV. EVALUATION

To verify the effects of the proposed system, we established a server environment with an Intel i7 Quad core CPU, 8GiB of DDR3 RAM, 4GiB of DRAM running CentOS 6.5 with Linux 3.10. The server consists of an Apache webserver, PHP scripting language, and MySQL database. Server-side operates the components with the PHP script language to support the web interface and the web storage for the users. Figure 6 shows how the client-side consists of three clients, such as a server-based web client, desktop client, and a mobile client. At that time, we used the network protocol of the open source project, called *Owncloud*, to maintain the openness.

*A. Scenario for evaluation*

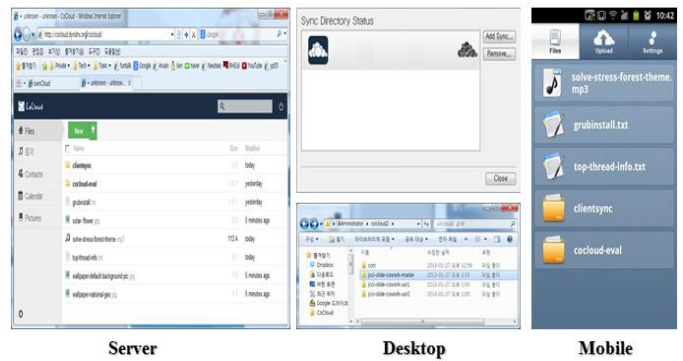

**Figure 6. The client applications for contents cooperation among users**

We evaluated how the proposed system automatically merges the content without causing any the *file conflict* for the shared file when users modify the content of the shared cloud file at the same time,

*1)* Cloud server: manages the cloud storage. [18] When the user changes the content of a file, the service daemon of the server executes the transaction of the data using a spin lock [19] mechanism and a FIFO queue.

*2)* Desktop software: client for a user's desktop computer. When the user modifies the shared cloud file and saves the modified file, the client requests the data synchronization of the cloud file to the cloud server.

*3)* Mobile software: allows mobile users to modify the shared file. At this time, the operation flow for the mobile program is similar to that of the client software.

### B. Experimental result: Revision History of Cloud File

Figure 7 shows how the modified content is saved into the distributed history management directory when many users execute the *check-in* command to merge modified content to the file. We can manage the modified content for an offline meeting by preserving the history of modified information among the users from many devices.

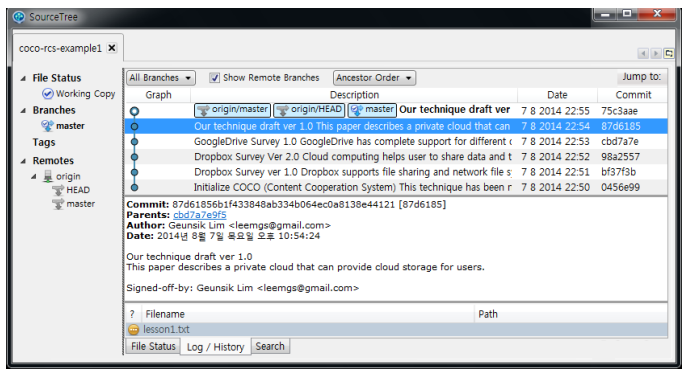

**Figure 7. the content revision history of the shared cloud file among the users**

### C. Experimental result: Single-user and multi-devices

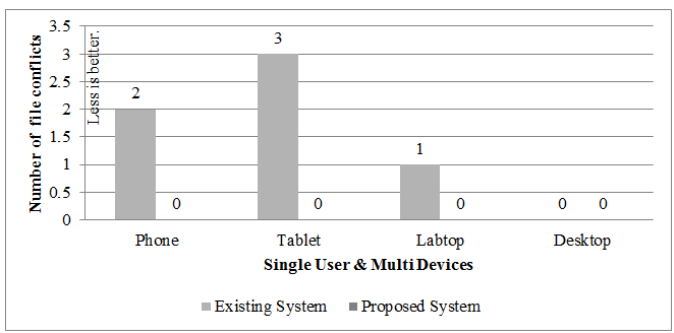

**Figure 8. File conflicts in single-user and multi-devices**

Figure 8 shows the experimental result of a *file conflict* from a single-user and a multi-device environment. The proposed system guarantees zero *file conflict* even when a user tries to modify a shared file from multiple devices simultaneously, as compared to an existing system. In Figure 8, the x-axis refers to the name of the device of the user. The y-axis refers to the number of *file conflicts* when the user tries to modify a shared file using multiple devices at the same time. The data synchronization mechanism of the existing system cannot solve *file conflicts* when the user tries to modify a shared file over two devices.

### D. Experimental result: Multi-users and single-device

Figure 9 shows a comparison between the existing system and the proposed system in terms of a multi-user single-device environment. We evaluated the performance of our proposed system with respect to the data synchronization of the shared cloud file without any *file conflicts* for the case where many users modify the shared file at the same time. The x-axis refers to the number of users that access to the shared file simultaneously. The y-axis refers to the number of *file conflicts* of the shared file. From our experiment, we have zero *file conflict* resulting from the automatic merging operations when many users modify the content of the shared file from anywhere at the same time.

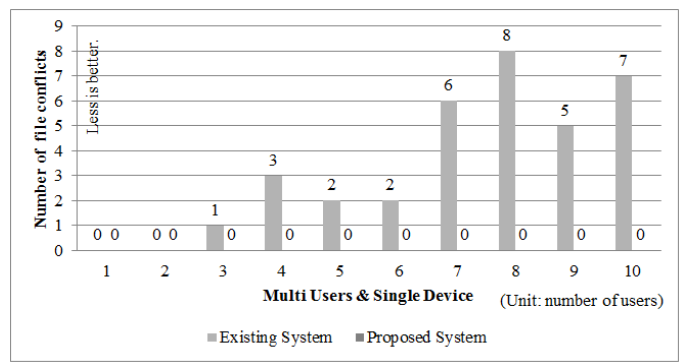

**Figure 9. File conflicts in multi-users and single device**

### E. Experimental result: Multi-users and multi-devices

Figure 10 shows a comparison between the existing system and the proposed system in terms of a multi-user and a multi-device environment. The proposed system recognizes multiple users with multiple devices better than the existing system, without any *file conflicts*, in order to support a collaborative educational system.

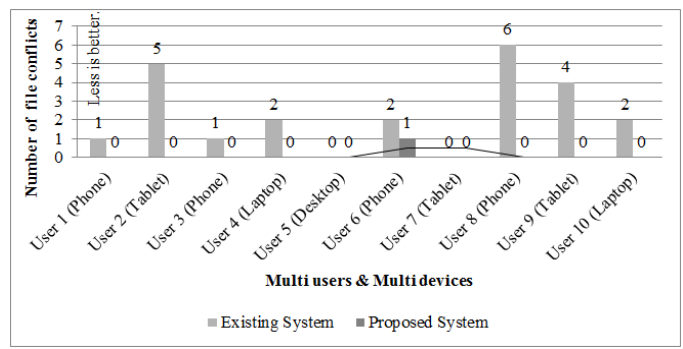

**Figure 10. File conflicts in multi-users and multi-devices**

The experimental results of the proposed system generated one file conflict even though 10 users accessed the shared file at the same time. Our analysis indicates that this problem happens because the user did not execute the *check-out* command after setting a manual lock preference without an automatic locking mechanism in order to merge the modified content manually.

*F. The difference comparison between the existing system and the proposed system*

Table 1 summarizes the strong points of the proposed system against those of the existing system. The proposed system synchronizes modified content units in order, being aware of multiple users with multiple devices. Moreover, it supports a history management technique for the modified cloud file in the case that too many users try to modify the shared file simultaneously. The proposed system solves the problem of an increased server management cost due to running a combined cloud storage and source code management system. We described the management technique for the file history for the case where many users modify the shared cloud file based on the distributed system structure.

**Table 1. Functional comparison between the existing system and our proposed system**

| Content | The Existing System | The Proposed System |
|---|---|---|
| Web-hard | Yes | Yes |
| Network File Sync | Yes | Yes |
| SCM Method | Modified-file based | Changed-content based |
| SCM Server Cost | Yes (Need additional server because there is not data locking mechanism) | No (Doesn't need additional server because cloud server is SCM server) |
| SCM Multi-user and SCM Multi-device | No (file conflict occurs) | FIFO-based spin-locking mechanism |
| SCM Linkage | No | Sync function of cloud client (Preference: Manual, Automatic) |
| SCM Merging | Coarse-grained method (file-unit based) | Fine-grained method (check-in time based) |

## I. Conclusion

The proliferation of smart devices, such as smart phones, smart tablets, and smart watches has quickly popularized mobile cloud services. Existing online education has moved from desktop environment to mobile due to the proliferation of smart devices. Therefore, it is important for many users to be able to edit content of cloud files in online environment from anywhere to pass the collaboration limitation, which *Dropbox* doesn't have.

This paper proposes a cloud-based content cooperation system that can modify cloud files in multi-user multi-device environments. It consists of a modified content-based cloud file locking mechanism, distributed file history tracking, automatic *check-in* and *check-out* method, and automatic revision control components. Moreover, we addressed the automatic merging of changes in the content of a cloud file by using the modified content unit. This mechanism helps many users to access shared cloud files using a FIFO-based spin-lock locking mechanism to avoid file conflict problem. In our experiment, we verified that the proposed system could merge shared files without causing any file conflict, and we compared our results against those of an existing system. We could browse the history of the shared file in order to track the task contents of each of the members.